# Flexibility-assisted heat removal in thin crystalline silicon solar cells

Seok Jun Han[a], Pauls Stradins[b], Sang M Han[a], Sang Eon Han[a*]

[a]Department of Chemical and Biological Engineering, University of New Mexico, Albuquerque, NM 87131, USA.

[b]National Center for Photovoltaics, National Renewable Energy Laboratory, Golden, CO 80401, USA.

*Corresponding author. E-mail: sehan@unm.edu

**Abstract:** Thin crystalline silicon solar photovoltaics holds great potential for reducing the module price by material saving and increasing the efficiency by reduced bulk recombination loss. However, the module efficiency decreases rather sensitively as the module temperature rises under sunlight. Effective, inexpensive approach to cooling modules would accelerate large-scale market adoption of thin crystalline silicon photovoltaics. For effective cooling, we exploit high flexibility of single-crystalline thin silicon films to create wavy solar cells. These wavy cells possess larger surface area than conventional flat cells, while occupying the same projected area. We experimentally demonstrate that the temperature of thin wavy crystalline silicon solar cells under the sunlight can be significantly reduced by increased convective cooling due to their large surface area. The substantial efficiency gain, achieved by the effective heat removal, points to high-performance thin crystalline silicon photovoltaic systems that are radically different in configuration from conventional systems.

## 1. Introduction

Solar photovoltaics (PV) market is growing rapidly on a global scale. To further accelerate this trajectory, the desired goal is to enhance the efficiency and reduce the cost of PV systems. Among numerous approaches that have been explored to achieve this goal, one economically viable direction is to use thin (<50 μm) crystalline silicon (c-Si) substrates [1-7]. The economic benefit of using thin c-Si substrates is three-fold. First, today's manufacturing processes to produce thin c-Si substrates can minimize material loss. Conventional sawing methods to produce a few 100-μm-thick c-Si wafers typically waste an excess of 100 μm per wafer [8]. In contrast, thin c-Si substrates can be prepared by various loss-minimizing techniques, such as epitaxial growth on porous Si [9], exfoliation from a thick c-Si substrate [10], and liquid-phase crystallization of amorphous Si [11]. Second, thin c-Si substrates can reduce the material usage and ultimately module cost. Current c-Si solar cells typically use 100 to 300-μm-thick Si wafers. Because thick c-Si wafers comprise 20-30% of a typical solar module cost [12], the use of thin c-Si substrates can translate to a substantial reduction in the module cost. Third, thin c-Si substrates would reduce the module weight and therefore transportation and installation cost. The PV industry is currently moving forward with bifacial double glass construction using heavy 2.5-mm-thick glass panels. In comparison, the thin c-Si solar cells can be highly flexible and would eliminate the design constraint of having rigid glass panels. The flexible c-Si cells can be supported on a light-weight platform, opening up a possibility to reduce the balance of systems cost.

Thin c-Si cells, however, do not absorb sunlight sufficiently and need to be complemented by efficient light-trapping features to enhance photovoltaic efficiency. Much effort has been devoted to developing effective light-trapping structures to improve the efficiency, including periodic [13, 14], quasi-random [15, 16], and random structures [17, 18]. While quasi-random, or deterministically random, structures can produce extremely efficient light absorption, they pose great challenges for cost-effective manufacturing. Random structures are currently the most affordable choice in industry. In comparison, periodic structures make the device manufacturing more controllable and provide highly efficient light trapping. Based on these advantages, various inexpensive, wafer-scale lithography techniques are being developed for fabricating periodic structures on c-Si surfaces [19, 20]. The record efficiency (15.7 %) solar cell for a 10-μm-thick c-Si substrate made use of a periodic array of nanoscale pyramidal etch pits, known as inverted nanopyramids, on the front surface [1]. We have predicted that the efficiency of such thin c-Si cells can be further enhanced by 0.9-1.9 % by breaking the point group symmetry of the inverted nanopyramid array without using expensive off-cut silicon wafers [14].

As the thin c-Si solar cells become increasingly absorptive using efficient light-trapping features, the cell temperature is likely to increase progressively under the sunlight. High temperatures, in turn, reduce the photovoltaic efficiency by various mechanisms, such as light- and elevated temperature-induced degradation [21, 22] that is not very well understood. The temperature-induced degradation is generally more pronounced for lower bandgap materials [23]. Among popular semiconductor materials for solar PV, c-Si has a relatively low bandgap. Consequently, the efficiency of c-Si solar cells is more sensitively affected by temperature than that of solar cells based on higher bandgap materials such as GaAs, CdTe, and CdS [23]. One way to counter this temperature susceptibility of c-Si cells is to cool the solar panels. A variety of cooling technologies for PV modules exist today [24-27], such as radiative cooling, water circulation, water sprinkling, immersion in water, thermoelectric cooling, air circulation, and phase-change materials. However, these technologies increase the manufacturing and/or operating cost, so that they are not commonly used in commercial products. For thin solar cells, a new cooling scheme can be considered, which exploits great flexibility of thin substrates to increase the efficiency. For example, we can think of a case where the maximum electric power generation is desired from the fixed area of a rooftop. Flexible solar cells can be shaped in a variety of 3D forms to generate a greater electric power than flat PV modules. Moreover, these 3D shapes have a larger surface area than a flat sheet, so that heat is more efficiently removed from the solar PV modules to the ambient air. Based on the high temperature sensitivity of c-Si PV modules, the 3D shaping would be highly beneficial for thin c-Si modules.

In this work, we fabricate thin, free-standing, flexible, 10 to 14-μm-thick c-Si solar cells and enhance their photovoltaic efficiency by both light trapping and flexibility-assisted heat removal. To enhance light trapping in inverted nanopyramid arrays, we employ a cost-effective method of breaking symmetry in the structures [14]. For efficient heat removal, we shape the thin c-Si cells in a wavy form. Based on the experiment, we predict substantial gains in photovoltaic efficiency from the thin wavy c-Si solar cells by making use of the large cell surface areas for heat removal.

## 2. Symmetry breaking

Symmetry breaking can enhance the overall absorptance of solar cells by increasing the number of resonant absorption peaks within the solar spectrum [14, 28, 29]. To experimentally determine the symmetry-breaking effect on photovoltaic efficiency, we fabricated 14-μm-thick free-standing c-Si cells with inverted nanopyramid arrays of $C_{4v}$ (Fig 1A) and $C_2$ (Fig 1B) symmetries on the front surface of the cells. For the proof-of-concept, thin c-Si substrates were prepared by etching thick *p*-type c-Si(100) wafers in a potassium hydroxide (KOH) solution [30] (Fig. S1). The inverted nanopyramid arrays were then introduced to the front surface of the thin substrate by interference lithography and subsequent etching in a KOH solution. Symmetry breaking was achieved by rotating a $C_{2v}$ symmetry etch mask around the c-Si [001] axis, such that a lattice vector in the mask is tilted from the [110] direction by 22.5° [14]. After the light-trapping structures were fabricated, we made an *n*-type emitter, using spin-on-dopant (P-1200, Desert Silicon) on the structured front surface. Boron was diffused into the substrate near the back surface, using a *p*-type spin-on-dopant (B-200, Desert Silicon), to create an Ohmic contact. Then, thin $SiO_x$ passivation layers were thermally grown on both front and back surfaces to a thickness of about 1.3 nm (Fig. S4). Intrinsic amorphous Si was deposited on the $SiO_x$ layers (Fig. S5) and subsequently turned into polycrystalline Si (poly-Si) by annealing at 950 °C for 3 hours. During the annealing process, the dopants in the c-Si substrate diffused out into the poly-Si through the $SiO_x$ passivation layers, forming highly doped poly-Si. Metal contacts were deposited by electron beam evaporation on the highly doped poly-Si (Fig. S6). The poly-Si films, each of which is 80-nm-thick, on the tunneling $SiO_x$ layers formed passivated contacts. After fabricating metal contacts, an 80-nm-thick $SiN_x$ antireflection layer with a refractive index of 1.9 was deposited on the front surface (Fig. S7). A schematic of the final structure is shown in Fig. 1D inset. Details of the fabrication process are available in Supplementary Material.

The $C_{4v}$ symmetry structure (Fig 1A) with a 700 nm × 700 nm unit cell has a four-fold rotational symmetry axis and four mirror-symmetry planes. In comparison, the $C_2$ symmetry structure (Fig 1B) has a 800 nm × 900 nm unit cell, characterized by a two-fold rotational symmetry axis and the absence of mirror symmetry. In both structures, the four edges at the base of the inverted nanopyramids correspond to [110] directions. Some parts of the (100) plane remained unetched, leaving flat areas. Further etching can remove the flat areas potentially increasing overall light absorptance [13, 14].

Figure 1C compares absorptance spectra of the two cells approximately above the bandgap of c-Si, which corresponds to $\lambda \leq 1.1$ μm. The symmetry breaking from $C_{4v}$ to $C_2$ increases absorptance both in the short and long wavelengths in the measured spectrum. The absorptance spectra in Fig. 1C are slightly different from those in our previous study [14] where, unlike the cells, both top metal electrode and poly-Si were absent. Absorptance for $\lambda \leq \sim 0.8$ μm is lower than that in our previous study [14] because of the top metal electrode which covers ~10 % of the c-Si surface (Fig. 1C inset). For $\lambda \geq \sim 0.95$ μm, absorptance is higher than that in our previous study [14]. The reason is due to the heavy doping and the poly-Si layers (Fig. 1D inset) as shown in Fig. S12.

The current-voltage (JV) characteristics of the cells with the different symmetry nanostructures are shown in Fig. 1D. Because of the increased absorptance by symmetry breaking, the short circuit current density $J_{sc}$ of the $C_2$ symmetry structure (29.9 mA/cm$^2$) is higher than that of the $C_{4v}$ symmetry structure (28.2 mA/cm$^2$) by 1.7 mA/cm$^2$, which corresponds to an enhancement of 6 %. The open circuit voltage $V_{oc}$ and the fill factor *FF* of the two cells are the same at 580 mV and 78%, respectively. The JV characteristics give efficiencies of 12.5 % and 13.6 % for the $C_{4v}$ and $C_2$ symmetry structures, respectively. The efficiency enhancement of 1.1 % by the symmetry breaking is slightly higher than the prediction of 0.9 % made in our previous study [14].

## 3. Wavy cell

Freestanding thin c-Si films are capable of being shaped in a variety of forms due to their high flexibility. For illustration, Fig. 2A shows a 14-μm-thick c-Si film that is bent with a radius of curvature down to less than 1 mm. Using the high flexibility, we shaped our thin c-Si cells into a wavy form using polymer templates. For the shaping, a 38-μm-thick aluminum (Al) film was pressed between two polymer templates with wavy surfaces. After removing the top polymer template, the conformally shaped Al film remained on the bottom template (Fig. 2B). Ethylene-vinyl acetate (EVA) was placed on the Al film and heated at 130 °C to be melted. A thin c-Si cell was placed on the EVA adhesive (Fig. 2C). Then, the top polymer template, which had previously been used to shape the Al film, was pressed onto the cell (Fig. 2D). The composite layers of templates, cell, EVA, and Al were cooled in the pressed state and the EVA was solidified. The two templates were subsequently removed to leave the cell glued on the Al film shaped into a wavy form of a periodicity of 4.8 mm (Fig. 2E). Surface area of the wavy structure was ~20 % greater than the projected area on the lateral plane. Similar wavy c-Si diode films were fabricated for stretchable electronics [31] and macroscopic shaping of solar cells can be carefully designed to enhance light trapping [32].

We fabricated a 9.5-μm-thick flat c-Si cell with the $C_2$ symmetry light-trapping structure and measured its IR absorptance spectra at an incident polar angle of $\theta = 8°$ and azimuthal angles of $\varphi = \pm 45°$ with the angles defined in Fig. 3A. The cell structure was the same as Fig. 1D inset except that the top metal contact was absent. Then, the cell was shaped in a wavy film, and the same absorptance measurement was performed on the wavy cell. The wave was periodic in the approximately diagonal direction of the cell in our experiment. Figure 3B compares the IR absorptance spectra of the cell before and after the shaping. While the polar angle $\theta = 8°$ is small, absorptance of the wavy cell shows a pronounced dependence on the azimuthal angle. Absorptance of the wavy cell is higher and lower than that of the flat cell at $\varphi = 45°$ and $-45°$, respectively. By controlling the cell orientation with respect to the wave direction, the overall absorptance could be maximized for practical applications.

To investigate the effect of the wavy form on the photovoltaic properties and heat removal, we fabricated a 14-μm-thick c-Si cell that has metal contacts and the $C_2$ symmetry light-trapping structure on the front surface. This cell gave an efficiency of 12.4 % which is lower than the efficiency 13.6 % of the cell with the $C_2$ symmetry structure in Fig. 1D, even though the cell thickness was the same. The lower efficiency may be due to some defects that were visibly detectable. Because the temperature dependence of JV characteristics for thin c-Si cells can be different between flat and wavy forms, after the photovoltaic properties of the flat cell were characterized as a function of temperature, the same cell was shaped in a wavy film and the photovoltaic characterization was performed on the wavy cell. For the temperature dependence, we varied the cell temperature from 25 °C to 70 °C as shown in Fig 4A. When the temperature was lowered back to 25 °C after heating, the JV curve returned to the original one.

At room temperature, the $J_{sc}$ in the wavy form (30.8 mA/cm$^2$) is lower than that in the flat form (31.3 mA/cm$^2$) by 1.6 % (Fig. 4A), indicating that the wavy shaping slightly decreased light absorption in the normal direction to the cell. The $V_{oc}$ and $FF$ have been negligibly affected by the wavy shape. As a result, the photovoltaic efficiency $\eta$ in the wavy form (12.3 %) is slightly lower than that in the flat form (12.4 %) at room temperature.

As the temperature increases over our range of measurement, $V_{oc}$ decreases significantly and $J_{sc}$ increases slightly in both flat and wavy forms. Because of this dependence, $V_{oc}$ could be used as a good measure of the temperature. From the JV curves in Fig. 4A, we obtain temperature dependence of the photovoltaic efficiency as shown in Fig. 4B. In both flat and wavy forms, as the temperature increases, the efficiency decreases at a rate of –0.06 % / °C, which is equivalent to $(d\eta/dT)/\eta = -0.0048$ to $-0.0062$ / °C. This rate for the thin cell is similar to that for conventional thick cells [23]. The efficiency of the cell is slightly higher in the flat form than in the wavy form by ~0.2 % at each temperature due to a higher $J_{sc}$.

The wavy cell has a surface area ($S_w$) larger than the projected area ($S_0$) on its lateral plane by a factor of $\alpha \equiv S_w / S_0 = 1.2$. The increased surface area would enhance heat removal from the cell when it is heated by sunlight. The relation between $\alpha$ and the cell temperature can be found from a heat balance. Consider that a flat cell and a wavy cell are exposed to a solar flux $Q$ and the heat transfer from the cells to the ambient air is assumed to be described by a heat transfer coefficient $h$. The flat cell area is assumed to be the same as the projected area of the wavy cell on the lateral plane ($S_0$). Efficiencies of the flat ($\eta_f$) and wavy ($\eta_w$) cells depend on their temperatures $T_f$ and $T_w$, respectively. Solar average absorptance of the flat and wavy cells is $A_f$ and $A_w$, respectively. When radiative heat transfer through mid-IR is negligible, heat balance for the flat and wavy cells gives

$$2h(T_f - T_a) = [A_f - \eta_f(T_f)]Q \tag{1}$$

$$2h\alpha(T_w - T_a) = [A_w - \eta_w(T_w)]Q, \tag{2}$$

where $T_a$ is the ambient temperature. The factor of 2 on the left hand side of Eqs. (1) and (2) accounts for the top and bottom surfaces. From Eqs. (1) and (2), the ratio of the cell temperature differences from $T_a$ between the flat and wavy cells is obtained as

$$\frac{T_f - T_a}{T_w - T_a} = \left[\frac{A_f - \eta_f(T_f)}{A_w - \eta_w(T_w)}\right]\alpha. \tag{3}$$

If absorptance and efficiency of the flat cell are similar to those of the wavy cell, Eq. (3) becomes

$$\frac{T_f - T_a}{T_w - T_a} \cong \alpha > 1. \tag{4}$$

Accordingly, the wavy cell would be at a lower temperature than that of the flat cell. The lower temperature of the wavy cell would, in turn, lead to an efficiency gain.

To quantitatively assess the effect of the wavy form on the temperature and efficiency of the cell, we fabricated a flat cell that is identical to the wavy cell except its flatness as a reference. We measured the temperatures of the flat and wavy cells under the sunlight as shown in the left picture in Fig. 4C. The wavy cell was the same one that produced the results in Fig. 4A and B. The measurement was performed on July 9th and 21st, 2019 in Albuquerque under a clear sky. The cells were placed 15 – 17 mm above a polyethylene (PE) film with four pieces of spacers made of Styrofoam on the cell edges and were located ~1 m above the ground. We set up side walls in the measurement system with a PE film to minimize the effect of time-varying wind speeds on the cell temperature. In addition to the wind-blocking side walls, the wind speed was relatively low (~3 miles per hour) on the both dates of the experiments. Due to the low wind speed, the top of the measurement system was safely open, which allowed us to accurately measure the cell temperature with an IR camera from above. The cell temperature was obtained by calibrating the IR camera readings (Fig. S9) and represented by an average over 6 different points over the cell. The cells were tilted from the horizontal plane by 12° to face the sunlight in the normal direction. When the cell temperatures were measured at 1:30 pm, the zenith angle was 12°, the solar intensity was measured to be 1044 W/m², and the ambient temperature was 31 °C on the both dates. The measurements were performed with the cells at open circuit condition.

In the experiment on July 9th, 2019, the cells were on Al films and some parts of the Al surfaces were not covered by the cells. We did not cut the exposed Al parts because the cutting might break the cells if not carefully performed. The temperatures of the flat and wavy cells were 54 °C and 49 °C, respectively, so that the wavy cell was at 5 °C below the flat cell temperature. On July 21st, 2019, the same experiment was performed by painting the exposed Al surfaces black (Fig. 4D). The black paint increased the cell temperature in comparison with the Al surfaces, resulting in the flat and wavy cell temperatures of 69.7 °C and 64 °C, respectively (Fig. 4E). Thus, the wavy form of the cell decreased the cell temperature by 5.7 °C in the presence of the black paint.

The cell temperatures would have been different from the measurements if the exposed Al or black surfaces had been absent. To estimate the cell temperatures in this case, we obtain $h$ from the experiments. For our experiments, heat balance for the flat and wavy cells gives

$$2h(T_f - T_a) = <A_f> Q \qquad (5)$$

$$2h\alpha(T_w - T_a) = <A_w> Q, \qquad (6)$$

where $<>$ denotes an average over the area that includes both the cell and the exposed Al or black paint. Note that, in Eqs. (5) and (6), $\eta$ is absent unlike Eqs. (1) and (2) because open circuit condition is considered. From Eqs. (5) and (6), we find that $h$ = 10.7 W/(m²K) and 10.2 W/(m²K) on the July 9th and 21st, respectively. These values are modestly higher than $h \sim$ 8 W/(m²K) for natural convection [33]. When the exposed Al or black surfaces are absent, the cell temperatures are determined by replacing $<A_f>$ and $<A_w>$ by $A_f$ and $A_w$ in Eqs. (5) and (6), respectively. At open circuit condition without exposed Al or black paint, we obtain that $T_f$ = 63.9 °C and $T_w$ = 58.0 °C for the July 9th experiment and that $T_f$ = 65.6 °C and $T_w$ = 59.3 °C for the July 21st experiment. On average, the temperature difference is $T_f - T_w$ = 6.1 °C at open circuit condition.

When an electric power is generated by the cells, the cell temperatures would be lower than those at open circuit condition. Substituting $\eta_f(T_f)$ and $\eta_w(T_w)$ obtained from Fig. 4B in Eqs. (1) and (2), at the maximum power generation condition, we find that $T_f$ = 58.8 °C and $T_w$ = 53.7 °C for the July 9th experiment and that $T_f$ = 60.3 °C and $T_w$ = 54.9 °C for the July 21st experiment. The average temperature difference is $T_f - T_w$ = 5.3 °C. The average efficiencies are obtained as $\eta_f(T_f)$ = 10.4 % and $\eta_w(T_w)$ = 10.5 %. Therefore, we estimate that, when the cell is shaped in the wavy form, the photovoltaic efficiency increases by 0.1 % by reducing the cell temperature by 5.3 °C.

Using Eq. (1) and Fig. 4B, we predict the temperature and efficiency of our 14-μm-thick c-Si cell at the maximum electric power generation condition as the area enhancement factor $\alpha$ of the cell changes by shaping in a nonflat form. We assume that the cell absorptance remains constant as $\alpha$ varies and the heat transfer coefficient is independent of the cell geometry. We consider that the cell is laminated between two polyethylene terephthalate (PET) films of a thickness 154 µm and a refractive index 1.575. For heat transfer calculations, a mid-infrared (IR) emissivity of the films is approximated to be 0.9 and an atmospheric transmittance is simulated by assuming 25 °C and 50 % relative humidity at an altitude of 1 km in MODTRAN [34]. Since our c-Si cells have a low mid-IR emissivity (note that they appear dark in Fig. 4E), the PET films lower the cell temperature significantly by radiative cooling [35, 36]. For example, compared to an absence of mid-IR radiative heat transfer, the PET films decrease the flat cell temperature by 17 °C on a wind-free day under our simulation conditions.

Figure 5 shows our predictions for the AM1.5G solar spectrum when the ambient temperature is 25 °C. As $\alpha$ increases from 1 to 5, the cell temperature decreases by 14.2 °C at $h = 8$ W/(m$^2$K) (Fig. 5A). As a result, the efficiency increases significantly by 0.9 % from 11.3 % to 12.2 % (Fig. 5B). The temperature drop and the efficiency enhancement become less pronounced as $h$ increases. However, even at $h = 32$ W/(m$^2$K) which corresponds to a moderately high wind speed (~7 m/s) [37], the efficiency increases by a non-negligible amount of 0.3 % as $\alpha$ increases from 1 to 5. Note that, by increasing $\alpha$ from 1 to 4 at $h = 8$ W/(m$^2$K), we have an even greater efficiency than by increasing $h$ from 8 W/(m$^2$K) to 32 W/(m$^2$K) at $\alpha = 1$ (Fig. 5B). Therefore, increasing the cell surface area by 4 times in a nonflat shape is equivalent to increasing the wind speed by ~7 m/s in terms of the efficiency gain.

## 4. Conclusion

Our study indicates that thin c-Si solar modules can be shaped in numerous nonflat forms, assisted by their flexibility, and can be effectively cooled, without extra cooling equipment or materials, simply by an enhanced surface area. For high efficiencies, symmetry-breaking nanostructures can be used for effective light trapping. We can imagine that the nonflat thin solar modules can be located at multiple heights such that sunlight that is reflected from a module at a lower place can be reflected again from the backside of another module at a higher place and absorbed by a third module. In this type of configuration, the nanoscale light-trapping structure and the macroscale nonflat geometry / configuration of the module can be simultaneously optimized, while nontrivial, for maximal sunlight utilization. In general, the capability of thin c-Si solar cells in increasing their efficiency by effective and inexpensive heat removal may push the current configuration of photovoltaic systems into a radically different one.

**Acknowledgments**

The authors acknowledge financial support by grants DMR-1555290 (S.E.H.), CMMI-1635334 (S.E.H. and S.M.H.), and CHE-1231046 (S.M.H.) from the National Science Foundation. This work was authored in part by the National Renewable Energy Laboratory, operated by Alliance for Sustainable Energy, LLC, for the US Department of Energy (DOE) under Contract No. DE-AC36-08GO28308. Funding was provided by US Department of Energy Office of Energy Efficiency and Renewable Energy Solar Energy Technologies Office under agreements 30301 and 34359. The views expressed in the article do not necessarily represent the views of the DOE or the U.S. government. The U.S. government retains and the publisher, by accepting the article for publication, acknowledges that the U.S. government retains a nonexclusive, paid-up, irrevocable, worldwide license to publish or reproduce the published form of this work, or allow others to do so, for U.S. government purposes.

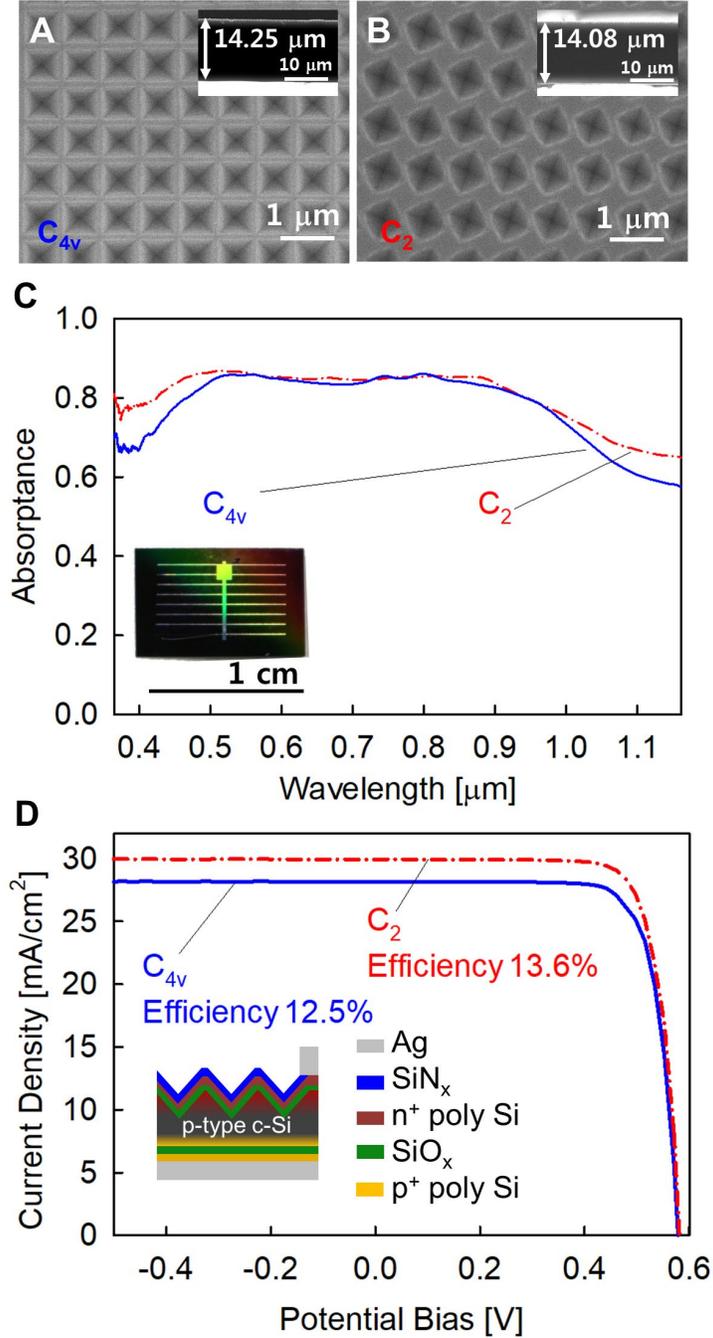

**Fig. 1.** Efficiency enhancement in thin c-Si solar cells by breaking nanostructure symmetry. Scanning electron micrographs of a top view of the (A) $C_{4v}$ and (B) $C_2$ symmetry inverted nanopyramid arrays on 14-μm-thick c-Si films. Insets are side views of the films. (C) Absorptance spectra of the two solar cells with the nanostructures of different symmetry. Inset is a picture of the solar cell with the $C_{4v}$ symmetry structure. (D) JV characteristics of the two solar cells. Inset is a schematic side view of the solar cells.

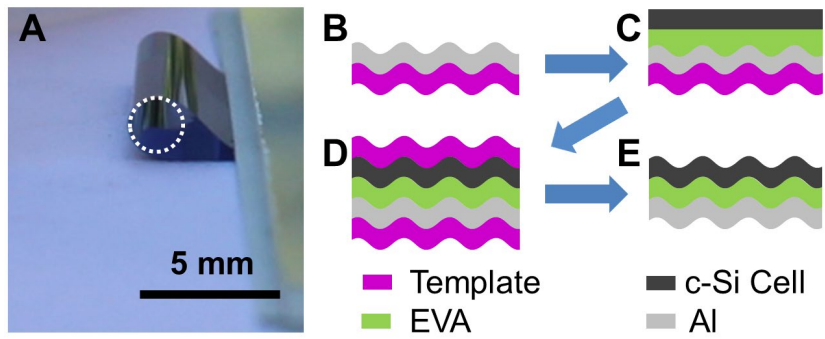

**Fig. 2.** High flexibility of thin c-Si solar cells. (A) A picture of a 14-μm-thick c-Si film bent with a radius of curvature less than 1 mm. (B to E) Illustration of the process for shaping thin c-Si solar cells into a wavy film using a template. The solar cells are glued on a wavy Al film by EVA.

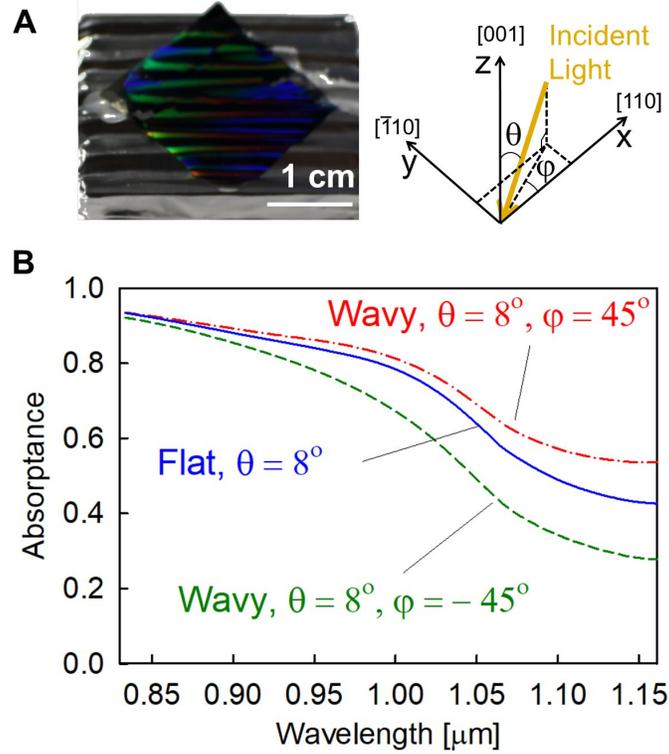

**Fig. 3.** Optical property of a wavy solar cell. (A) A picture of our 9.5-μm-thick wavy solar cell with the $C_2$ symmetry light-trapping structure. Top metal contact is absent and the solar cell is glued on an Al film (left). The solar cell orientation and incident light direction (right). (B) Absorptance spectra of the solar cell in (A) before and after being shaped into the wavy film.

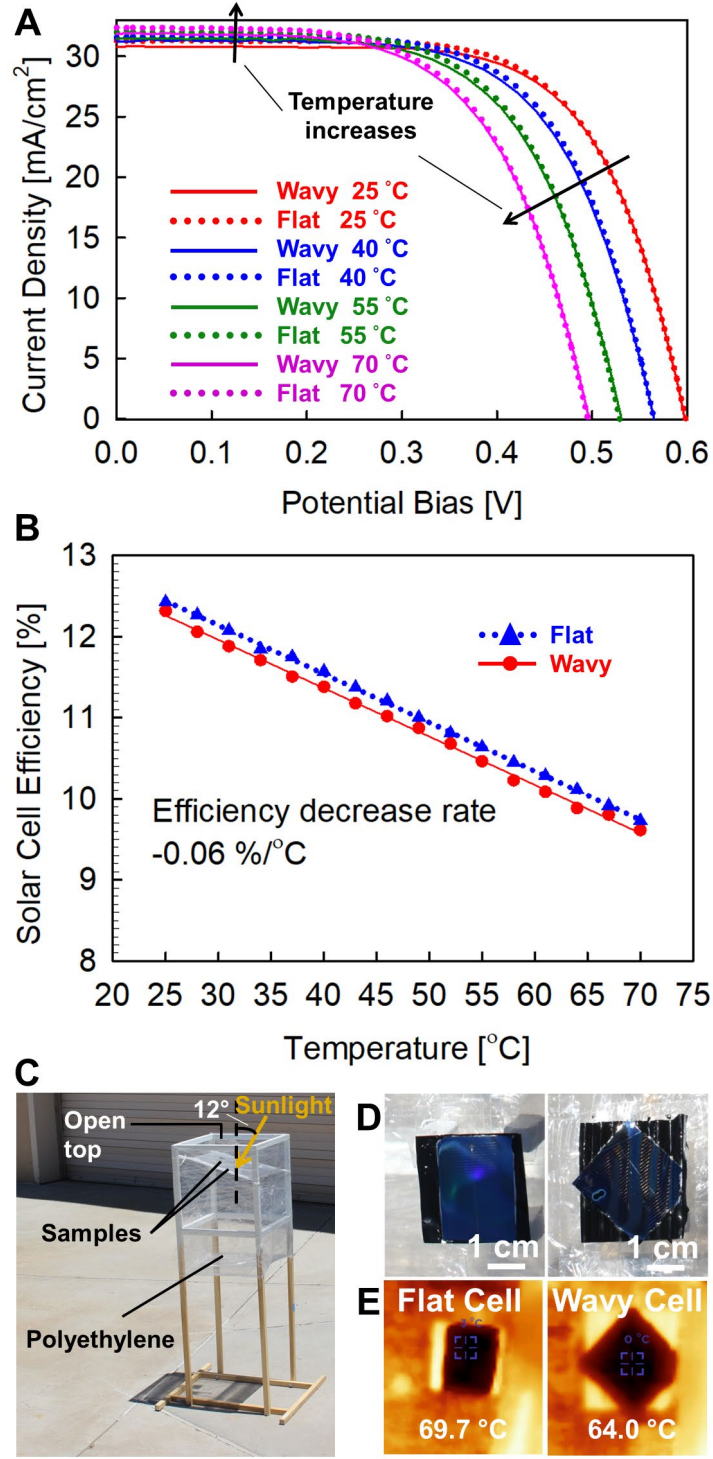

**Fig. 4.** Efficiency enhancement of a wavy solar cell by cooling. Temperature dependence of (A) the JV characteristics and (B) the photovoltaic efficiency of our 14-μm-thick c-Si solar cell with the $C_2$ symmetry light-trapping structure in both flat and wavy forms. (C) Our measurement setup for solar cell temperature under sunlight. (D) Optical and (E) IR images of our flat and wavy cells under sunlight.

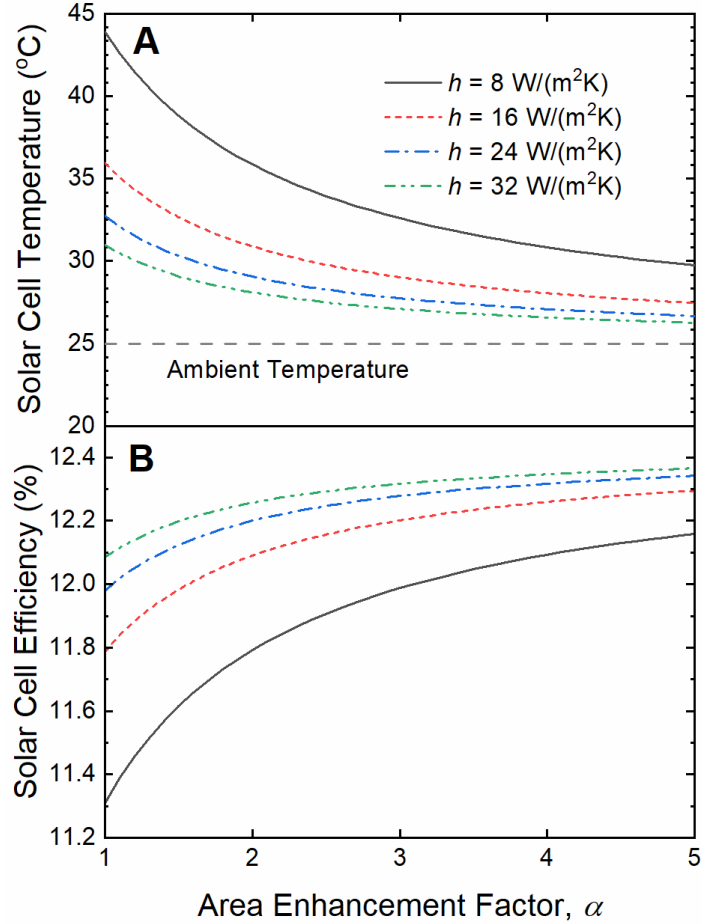

**Fig. 5.** Efficiency enhancement and cooling of thin c-Si solar cells by surface area enhancement. Predicted (A) temperature and (B) efficiency of our 14-μm-thick c-Si solar cell with the $C_2$ symmetry light-trapping structure as a function of the area enhancement factor for various heat transfer coefficient values at an ambient temperature of 25 °C.

# Supplementary Material for

# Flexibility-assisted heat removal in thin crystalline silicon solar cells

Seok Jun Han[a], Pauls Stradins[b], Sang M Han[a], Sang Eon Han[a*]

[a]Department of Chemical and Biological Engineering, University of New Mexico, Albuquerque, NM 87131, USA.

[b]National Center for Photovoltaics, National Renewable Energy Laboratory, Golden, CO 80401, USA.

*Corresponding author.  E-mail: sehan@unm.edu

Free-Standing Thin c-Si Film Preparation

We prepared free-standing thin (10-14 µm) c-Si films by etching 4-inch double-side-polished 250-µm-thick wafers in a KOH solution. The thick c-Si(100) float-zone p-type wafers were boron-doped and had a resistivity of 1-3 Ωcm. We performed etching of the wafers in a semiconductor-grade KOH solution of 50 wt% at 90 °C. To prevent contamination during etching, clean quartz wares were used for beakers and wafer holders. At the etching conditions, the wafer thinning rate was measured to be ~60 µm/hour/side as shown in Fig. S1.

To achieve uniform etching over a wafer, temperature gradient in the KOH solution was minimized by stirring the solution during etching. As hydrogen bubbles were vigorously generated as an etch byproduct, strong convective flows occurred in the solution, which would help maintaining etch uniformity. Thickness variation over a 4-inch wafer was negligible after thinning. Mirror-like clear surfaces without noticeable haze in the wafer were preserved after thinning [Fig. S2A and Fig. S2C]. Our surface profiling (Dektak 3, Veeco) revealed that local surface roughness increased from <10 nm to 20 – 60 nm by wafer thinning [Fig. S2B and Fig. S2D]. The surface roughness of the thinned wafers was sufficiently small for patterning with our interference lithography (IL) that uses a wavelength of 355 nm.

The free-standing thin c-Si films had to be handled carefully to avoid breakage. However, at a thickness of 10-14 µm, the c-Si films were able to be processed in our solar cell fabrication steps.

Light-Trapping Structure Fabrication

We fabricated light-trapping structures of $C_{4v}$ and $C_2$ symmetry inverted nanopyramid arrays on the thin c-Si films by IL and KOH etching [1]. For IL, a silicon nitride ($SiN_x$), an antireflection coating, and a photoresist were deposited in series on the thin c-Si films. The $SiN_x$ layer was deposited by plasma enhanced chemical vapor deposition (PECVD) at a thickness of 30 nm and a refractive index of 1.9 at 633 nm. In the PECVD, the ratio of volumetric flow rate between a diluted silane gas (5 vol% $SiH_4$ in Ar) and a pure ammonia gas was 4.0, and a deposition pressure was 1000 mTorr. The antireflection layer (iCON-16, Brewer Science) was spin-coated at a thickness of 160 nm. Then, a 500-nm-thick negative photoresist (NR7-500P, Futurrex) was spin-coated as a top layer. The photoresist layer was preheated for soft baking (150 °C, 60 seconds), patterned in Lloyd's mirror interferometry with a 355-nm YAG-Nd laser (Infinity 40-100, Coherent Inc.), and post-baked (100 °C, 60 seconds). After the photoresist was developed, reactive ion etching was performed with $O_2$ and $CHF_3/O_2$ for etching the antireflection and $SiN_x$ layers, respectively, to expose the c-Si surfaces through the etch windows. The exposed surfaces were anisotropically etched in a 20 wt% KOH solution at 50 °C for 10 minutes. In the solution, the photoresist and antireflection layers were delaminated, and inverted nanopyramids were etched into the c-Si surfaces with the $SiN_x$ as an etch mask. The $SiN_x$ was subsequently removed in a buffered oxide etch solution.

Thin c-Si Solar Cell Fabrication

Our thin solar cell fabrication process is summarized in Fig. S3. We fabricated thin solar cells based on float-zone p-type c-Si films. Inverted nanopyramid structures were introduced into the top surface of the film (Fig. S3A). An n-type (P-1200, Desert Silicon) and p-type (B-200, Desert Silicon) spin-on-dopant (SOD) were deposited on the front and back surfaces, respectively (Fig. S3B). The deposited SODs were diffused into the vicinity of the surfaces at 960 °C for 2 hours (Fig. S3C). After this predeposition, the surfaces were cleaned (Fig. S3D) and tunneling $SiO_x$ layers were thermally grown at 700 °C for 5 minutes on the surfaces for the passivation (Fig. S3E).  Because of the inverted nanopyramid structures, the $SiO_x$ layers were grown mostly on c-Si (111) planes on the front surface.  To determine the $SiO_x$ layer thickness, we grew an $SiO_x$ layer on a c-Si(111) wafer at the same heating conditions and measured the $SiO_x$ layer thickness.  In our ellipsometry measurement (M-2000, J. A. Woollam), we set the thickness of a c-Si/$SiO_2$ transition layer at 1 nm and included this layer as a film of an effective refractive index in estimating the thickness of the $SiO_x$ layer (Fig. S4 inset).  The $SiO_x$ layer thickness on the (111) surface was determined to be ~1.3 nm at our thermal growth condition (Fig. S4).  Then, to eliminate dangling bonds that potentially exist between the $SiO_x$ / c-Si interfaces, the c-Si film was treated in a hydrogen gas at 400 °C for 10 minutes.

After the hydrogen treatment, intrinsic amorphous Si (i-aSi) layers were deposited by plasma-enhanced chemical vapor deposition (PECVD) on the $SiO_x$ surfaces at a thickness of 80 nm (Fig. S3F).  In this step, it was important for high efficiency solar cells to obtain high quality i-aSi layers that were free from particles [2, 3].  To determine a deposition condition that prevents particles formation, we deposited relatively thick (4 – 8 μm) i-aSi films by varying the radiofrequency (RF) power and the pressure of a 5% silane gas in Ar ($P_{SiH4/Ar}$).  At various deposition conditions, we measured the mean diameter ($d$) and number density ($\rho$) of particles on the deposited i-aSi films.  Figure S5 shows optical microscope images of the deposited i-aSi layers at example deposition conditions.  From our experiment, we determined that PECVD at $P_{SiH4/Ar}$ = 450 mTorr and an RF power of 50 W resulted in high quality i-aSi layers [Fig. S5C].

To protect the i-aSi layers from contamination in a later step of dopant drive-in, we deposited 100-nm-thick $SiO_x$ layers on the i-aSi surfaces by PECVD.  This step can be skipped if the fabrication facilities are contamination-free.  After the protective layer deposition, the c-Si film was heated from 200 °C to 950 °C at a rate of 83 °C/hour, and maintained at the constant temperature for 3 hours to drive-in the dopants both into the deeper c-Si regions and into the i-aSi layers through the tunneling $SiO_x$ layers (Fig. S3G).  During this process, the i-aSi was simultaneously crystallized into polycrystalline Si.  The polycrystalline Si layers were highly doped to make good Ohmic contacts with metal electrodes.

After the protective $SiO_x$ layer was removed, metal electrodes were fabricated (Fig. S3H). For top metal contact, a c-Si mask with grid openings was fabricated by laser grooving and subsequent KOH etching.  Figure S6 shows a magnified optical image of the mask near a contact pad.  Top contact resulting from e-beam metal evaporation on a cell covered by the mask consisted of a contact pad, a 270-μm-wide bus bar, and metal fingers that are periodically placed in parallel to each other and perpendicular to the bus bar (Fig. 1C inset and Fig. S6).  The width of the metal fingers was 70 μm, and the spacing between neighboring fingers was 630 μm.  The top contact shaded about 10 % of the front surface of a cell, and was comprised of Ti, Ag, and Pd layers from bottom up with thicknesses of 20 nm, 1500 nm, and 20 nm, respectively.  Bottom metal contact (reflector) was fabricated by e-beam evaporation of Ag at a thickness of 1500 nm.

Finally, an 80-nm-thick $SiN_x$ antireflection layer was deposited on the front surface by PECVD at 300 °C (Fig. S3I).  Refractive index of the $SiN_x$ layer was measured by spectroscopic ellipsometry.  Figure S7 shows psi ($\psi$) and delta ($\Delta$) of our $SiN_x$ film and a $Si_3N_4$ film in literature [4], where $\tan\psi \cdot e^{i\Delta}$ is the ratio of reflection coefficients between p- and s-polarizations.  The good agreement in $\psi$ and $\Delta$ between the two films indicates that the composition of our $SiN_x$ film is similar to that of the film in the literature [4].  From $\psi$ and $\Delta$, we determined that the refractive index our $SiN_x$ film was 1.9 at 633 nm using the Cauchy model.

Contact Resistance Characterization

Contact resistance of our thin c-Si solar cells was characterized by transmission line measurement (TLM). For TLM, metal pads were fabricated on the top flat surface of our solar cells without a SiN$_x$ coating. The metal pads were of a size 627 μm × 2000 μm and the spacing between neighboring pads varied from 65 μm to 285 μm. The four-point-probe method was used to measure total resistance between pairs of metal pads. Figure S8 shows our measured total resistance as a function of the contact separation distance. From the graph and the width of metal pads, we calculated sheet resistance and contact resistance as 160.2 ohm/sq and 7 ohm, respectively. The transfer length, which is the mean distance that charge carriers travel beneath a contact pad before being collected by the pad, was calculated from the graph as 88.1 μm. From the transfer length, contact resistance, and pad width, the contact resistivity was obtained as 0.0125 ohm cm$^2$.

Optical Characterization

Optical absorptance of thin flat c-Si cells in the wavelength range of $\lambda$ = 0.4 μm – 0.9 μm was obtained using an integrating sphere (ISP-50-8R, Ocean Optics) coupled to a photodetector (USB4000-VIS-NIR, Ocean Optics). In the infrared (IR) range of 0.83 μm – 1.16 μm, absorptance spectrum was measured by a different spectrophotometer (VERTEX 70, Bruker) with an integrating sphere (A562, Bruker). For absorptance measurement of wavy cells, the cells needed be placed inside an integrating sphere and the Ocean Optics setup did not allow this configuration. Due to this equipment limitation, absorptance of wavy cells was measured only in the IR range using the Bruker setup (Fig. 3B).

Photovoltaic Characterization

Photovoltaic properties of the thin cells were characterized using a source meter (Keithley 2400) under an illumination of AM1.5G spectrum (ABET, LS-150). The illumination intensity of 1000 W/cm$^2$ was confirmed using a thermopile sensor (Newport, 919P-003-10). Voltage was varied from -0.5 V to 0.7 V with a step size of 0.018 V. The characterization system, which included four-point probes and a vacuum stage (TFI-5M, PV Measurements), was enclosed in a box and the temperature inside the box was controlled by convective heating. The temperature was increased from 25 °C to 70 °C with a step size of 3 °C. To ensure thermal equilibrium between the cells and the ambient air in the box at each photovoltaic measurement, the characterization was performed an hour after each target box temperature was reached.

Measurement of Solar Cell Temperature Under Sunlight

Temperature of thin cells under the sunlight was measured by an IR camera (RW-AAAX, Seek Thermal). Because the IR emissivity of the cells was much lower than a black body, the temperature readings of the IR camera were calibrated with actual temperatures determined by a thermocouple. Figure S9 shows the calibration results.

Solar Average Absorptance

In our heat transfer modeling, solar average absorptance of our flat thin c-Si solar cell with the C$_2$ symmetry structure was obtained from Fig. 1C. In the average, absorptance below the c-Si bandgap was set to zero. To estimate solar average absorptance of our wavy solar cell, we decreased the absorptance in Fig. 1C by 1.6 % which was the decrease amount in $J_{sc}$ as the solar cell was shaped in the wavy film (Fig. 4A). Solar average absorptance of the Al film and the black paint was obtained from Fig. S10A and B, respectively. The obtained absorptance values for the flat cell, wavy cell, Al, and black paint were 0.677, 0.665, 0.140, and 0.962, respectively.

Absorptance Difference Between Cell and Mono-Si Structure

The absorptance spectra in Fig. 1C are slightly different from those in our previous study [1]. The lower absorptance for $\lambda \leq$ ~0.8 μm in Fig. 1C than that in Ref. [1] is due to the top metal electrode. The reason for the higher absorptance for $\lambda \geq$ ~0.95 μm than that in Ref. [1] is not immediately clear. An indication for the reason is found in the fact that absorption coefficient $\alpha$ of heavily doped poly-Si in our cell is higher than monocrystalline Si by several orders of magnitude in near IR (Fig. S11). Thus, we compare the absorptance difference between our cell and Ref [1] for the $C_{4v}$ and $C_2$ symmetry structures with that for the Lambertian light-trapping limit. The comparison is shown in Fig. S12. In our dopant diffusion simulation [5], depth of a heavily doped region with a doping level of $1\times10^{20} - 8\times10^{20}$ cm$^{-3}$ is estimated to be ~200 nm. To account for top and bottom surfaces of the cell, we calculate the Lambertian limit for a thickness of 400 nm and take the difference between heavily doped poly-Si and monocrystalline Si. The inner cell region where heavy doping is absent is assumed to contribute negligibly to the absorptance difference. Figure S12 shows that the absorptance difference for both symmetry structures has a similar trend over the spectrum to that for the Lambertian limit. This implies that, if the cell electrodes were present for the Lambertian case, the 3 curves in Fig. S12 would roughly be similar to each other. More accurately, if we downshift the Lambertian limit curve by 0.15, assuming that absorptance decreases by this amount due to the electrodes, the 3 curves are similar to each other for $\lambda \leq$ ~0.8 μm but the Lambertian limit curve is higher than the others for $\lambda \geq$~0.8 μm. This agrees with the fact that our $C_{4v}$ and $C_2$ structures are slightly less efficient in light trapping than the Lambertian limit. Therefore, we conclude that the difference in the absorptance spectra in long wavelengths between Fig. 1C and Ref. [1] is due to the presence of heavy doping and poly-Si in our cell.

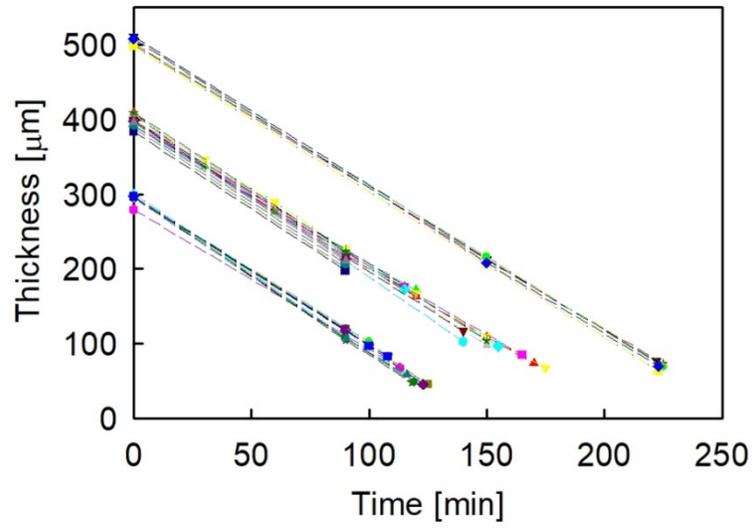

**Fig. S1.** Thickness of c-Si wafers as a function of etching time in a KOH solution of 50 wt% at 90 °C.

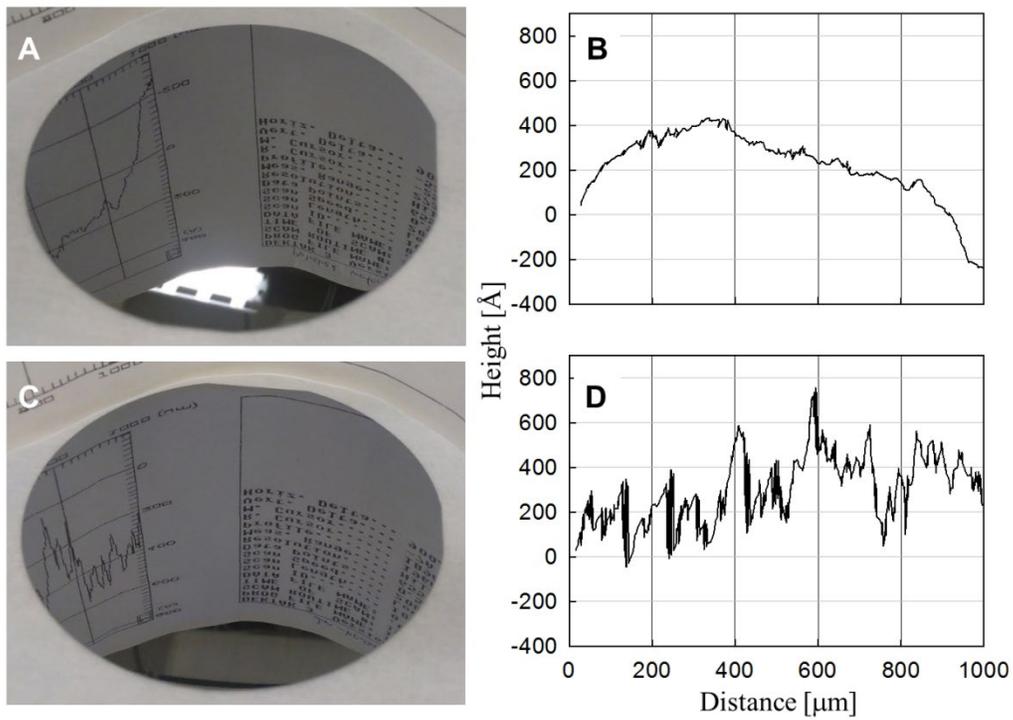

**Fig. S2.** Photo image and surface roughness profile of c-Si wafers (A,B) before and (C,D) after thinning by etching in a KOH solution.

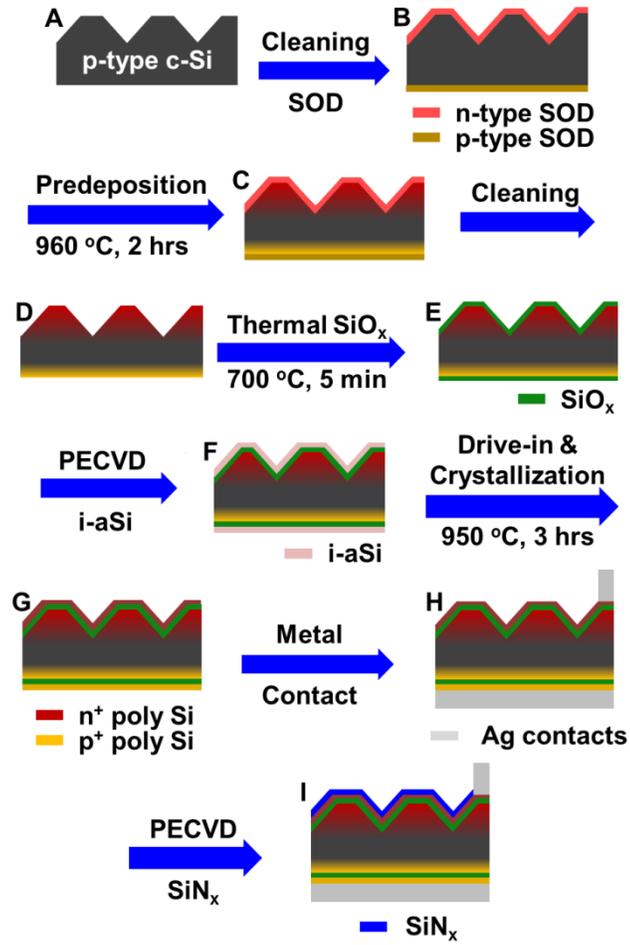

**Fig. S3.** Fabrication process flow for our thin c-Si solar cells.

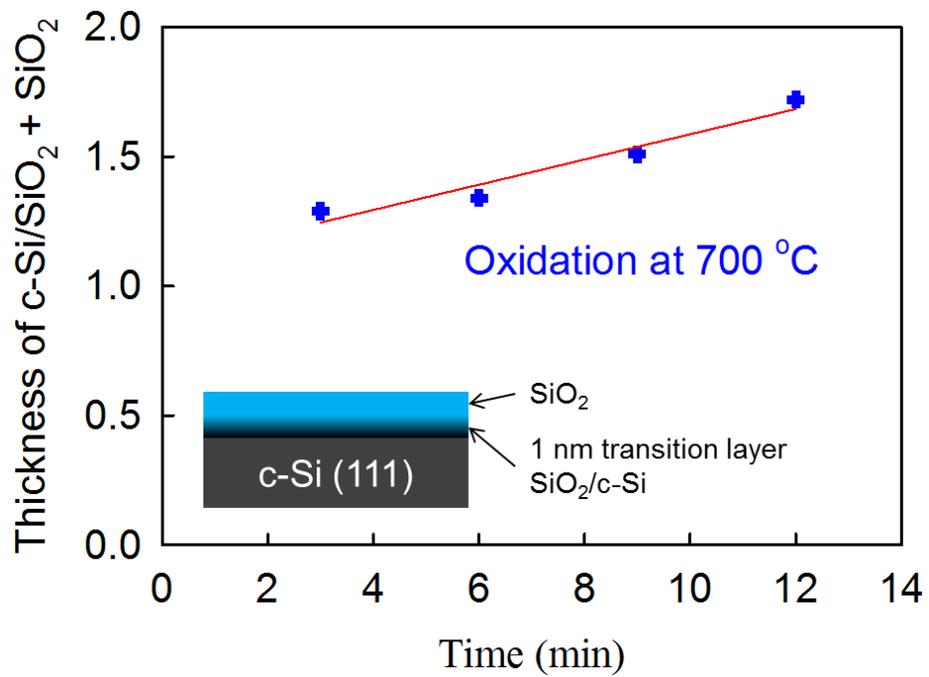

**Fig. S4.** Thickness of a SiO$_x$ layer thermally grown on a c-Si(111) surface at 700 °C as a function of time.

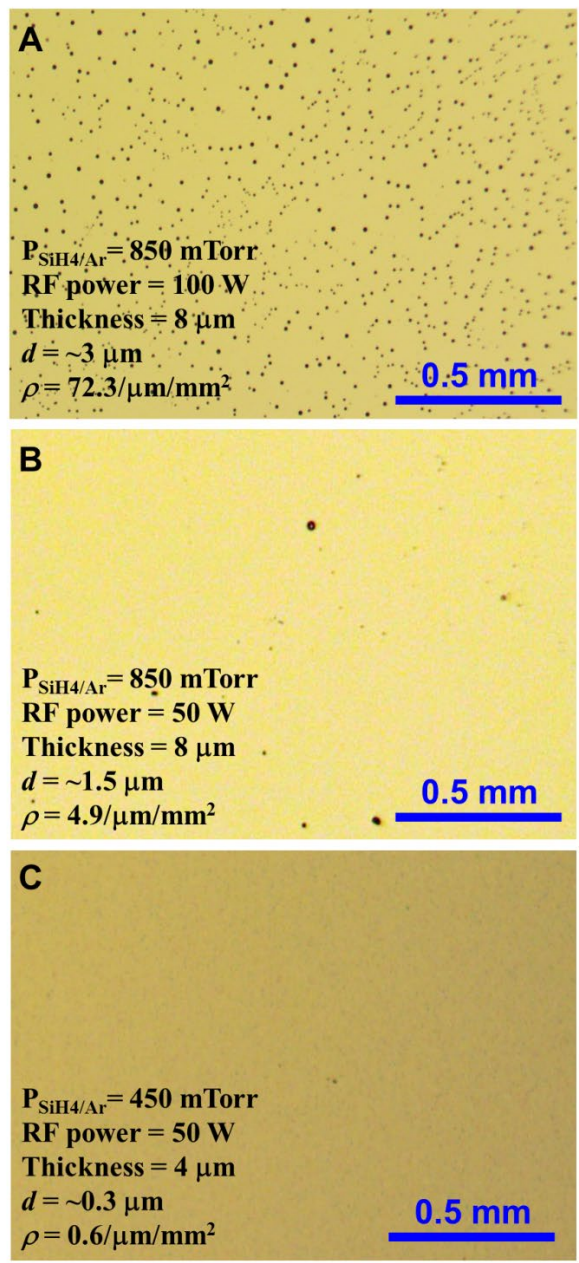

**Fig. S5.** Optical microscope images of intrinsic amorphous Si layers deposited by PECVD at 300 °C at various conditions.

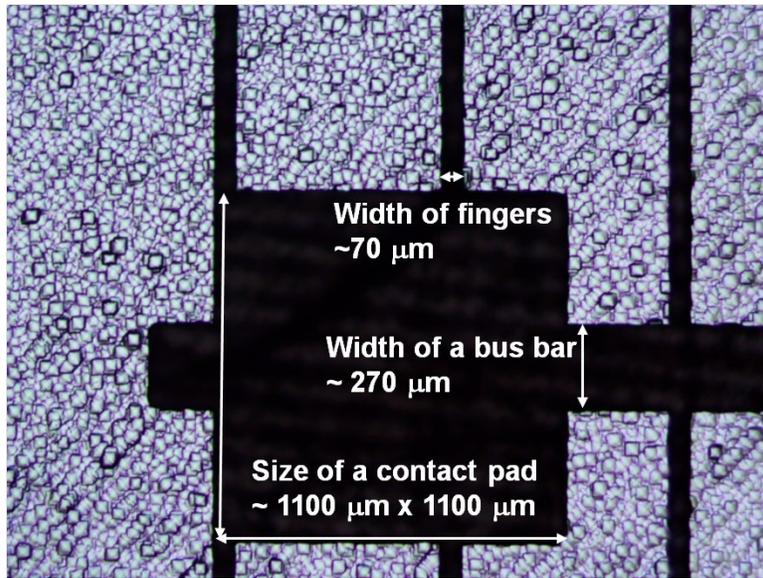
**Fig. S6.** Optical microscope image of our c-Si mask for top metal contact.

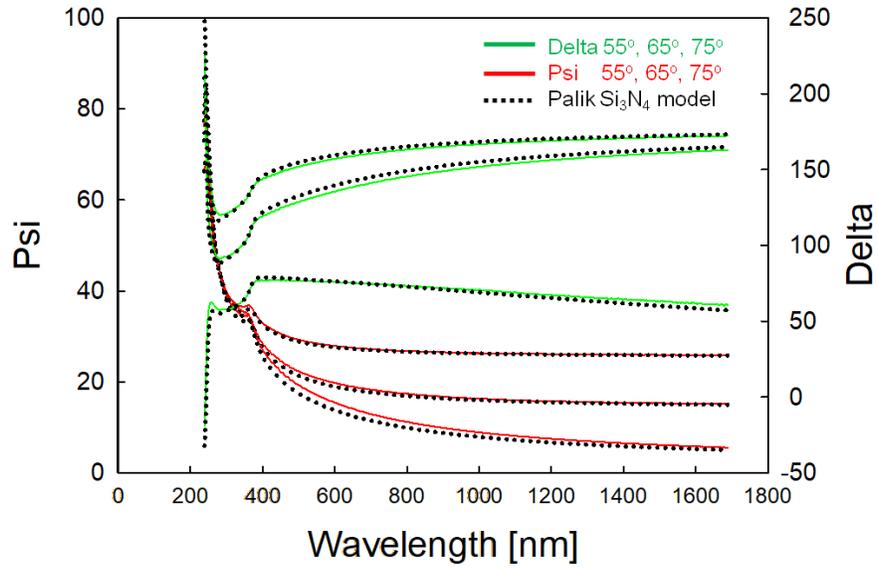

**Fig. S7.** Psi ($\psi$) and Delta ($\Delta$) in the fundamental equation of ellipsometry at various incidence angles for our $SiN_x$ film and a $Si_3N_4$ film in the handbook by Palik [4].

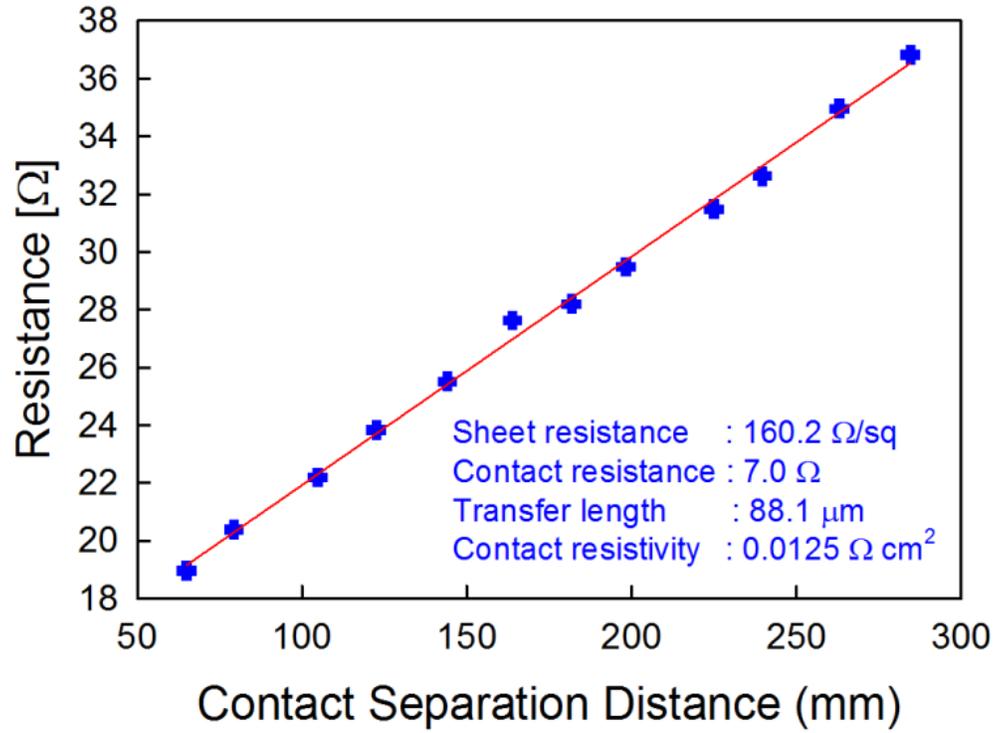

**Fig. S8.** Transmission line measurement results on our thin c-Si solar cells.

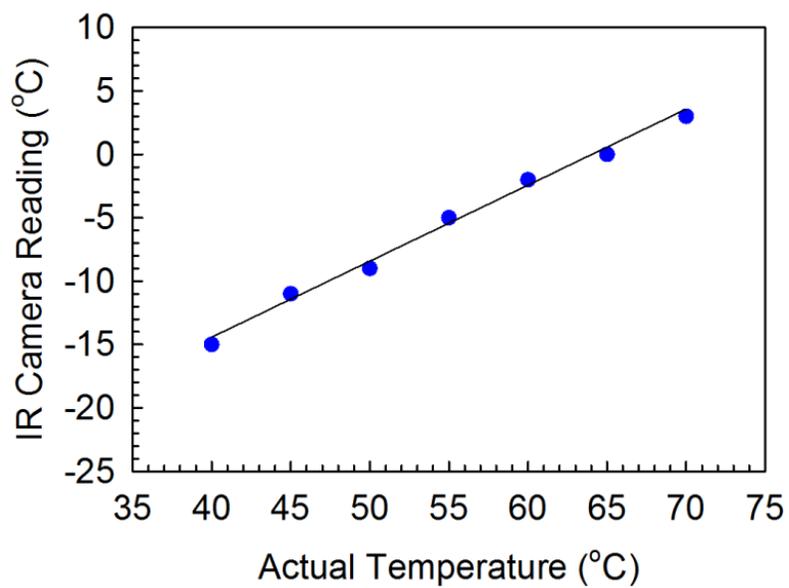

**Fig. S9.** Calibration line for IR camera readings on our thin c-Si solar cells.

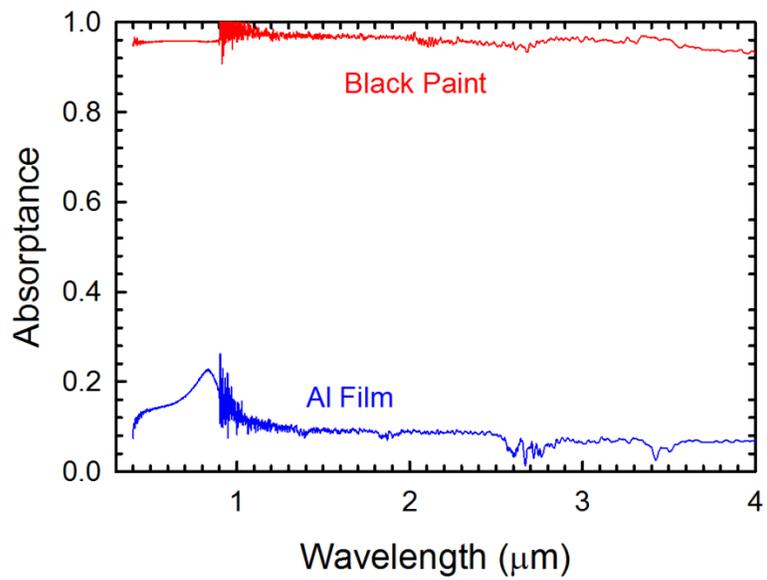

**Fig. S10.** Absorptance spectra of the Al film and the black paint used in our experiment.

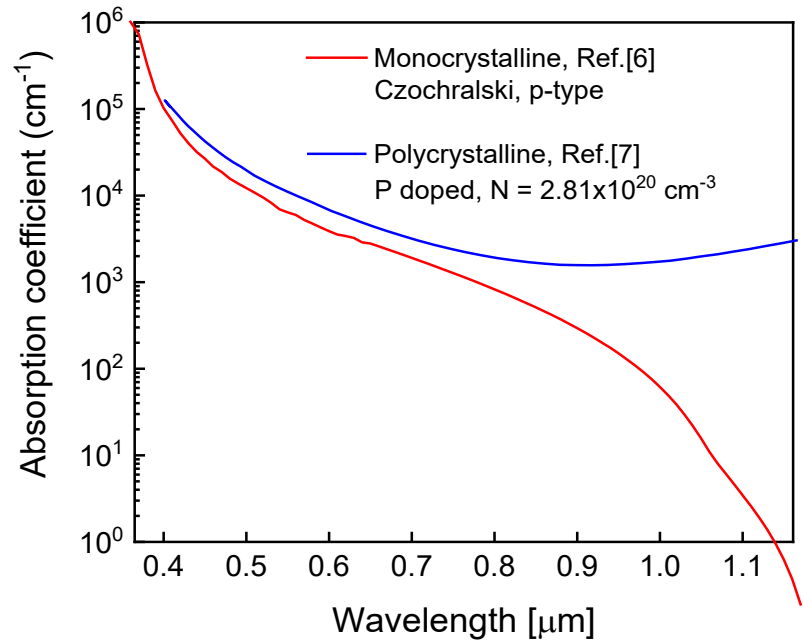

**Fig. S11.** Absorption coefficient for mono-Si [6] and heavily doped poly-Si [7] with phosphorus at a carrier concentration of $N = 2.81 \times 10^{20}$ cm$^{-3}$.

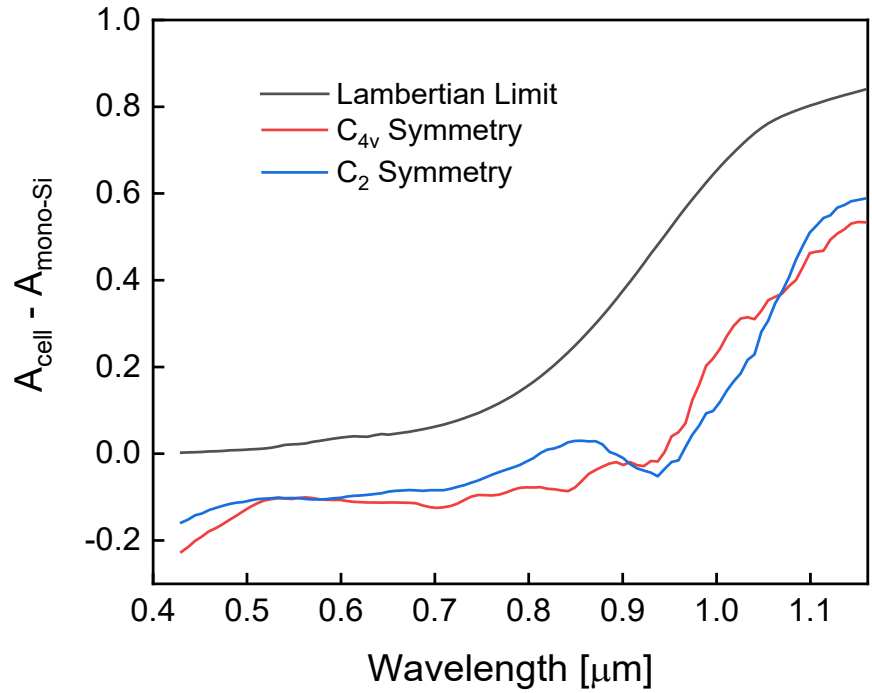

**Fig. S12.** Absorptance difference between our cell and the structure in Ref. [1]. For comparison, absorptance difference for the Lambertian light-trapping limit for a thickness of 400 nm is displayed. In the Lambertian limit calculation, absorption coefficients in Fig. S11 as well as the real part of refractive indices were used.